# Toward an Intent-Based and Ontology-Driven Autonomic Security Response in Security Orchestration Automation and Response

Zequan Huang[1,2], Jacques Robin[2], Nicolas Herbaut[1], Nourhène Ben Rabah[1], Bénédicte Le Grand[1]

[1] Center for Research in Informatics (CRI), University of Paris 1 Panthéon-Sorbonne.
[2] Engineering School of Informatics, Electronics and Automation (ESIEA), Paris.
{zequan.huang, jacques.robin}@esiea.fr, {nicolas.herbaut, nourhene.ben-rabah, benedicte.le-grand}@univ-paris1.fr

**Abstract.** Modern Security Orchestration, Automation, and Response (SOAR) platforms must rapidly adapt to continuously evolving cyber attacks. Intent-Based Networking has emerged as a promising paradigm for cyber attack mitigation through high-level declarative intents, which offer greater flexibility and persistency than procedural actions. In this paper, we bridge the gap between two active research directions – Intent-Based Cyber Defense and Autonomic Cyber Defense, by proposing a unified, ontology-driven security intent definition leveraging the MITRE-D3FEND cybersecurity ontology. We also propose a general two-tiered methodology for integrating such security intents into decision-theoretic Autonomic Cyber Defense systems, enabling hierarchical and context-aware automated response capabilities. The practicality of our approach is demonstrated through a concrete use case, showcasing its integration within next-generation Security Orchestration, Automation, and Response platforms.

## 1    Introduction

System and network infrastructures today face increasingly sophisticated threats, posing significant risks. Among the most challenging types of cyber threats are *Advanced Persistent Threats (APTs),* orchestrated by well-resourced attackers to gain prolonged unauthorized access to a target network to steal sensitive information or disrupt operations. From a defender's perspective, effective cyber defense encompasses three stages: (a) proactive prevention to deter threats before they occur, (b) real-time detection to identify breaches or anomalies, and (c) post-detection security responses to minimize damage – *i.e.*, attack mitigation. This paper concentrates on stage (c).

In APT scenarios, it is realistic to expect that skilled attackers can always find a way to evade the first line of defense, making real-time and post-detection response not just optional but essential. Modern *Security Operations Centers (SOCs)* still rely heavily on human experts to manually analyze and respond to incident alerts. However, human judgment errors are inevitable during mitigation operations. Despite advancements in **Security Orchestration, Automation, and Response (SOAR)** platforms, analysts still face overwhelming alert fatigue, due to the prevalence of false positives generated by insufficiently explained anomaly detection, leading to delayed or missed mitigation [1].

Current mitigation automation tools in SOAR, such as playbooks, operate as procedural checklists of steps and actions [2]. They are primarily concerned with enforcing low-level configurations, without inherently encoding the high-level security intents so far formulated by SOC analysts. Informally, a security intent indicates a desired system or network state in which the identified risk is mitigated. This gap – between operations and objectives – highlights the need for a higher-level security intent layer in security response automation, that aligns closely with human-understandable strategic goals. In other words, next generation SOAR platforms should derive and proceed with security intents "in mind", rather than relying on humans to bear these intents and explicitly instruct them on what attack mitigation actions to take. This transition represents a major step toward reducing human errors, reaction time, and sub-optimal trade-offs between conflicting security goals, such as availability, confidentiality, and integrity.

Research addressing security response automation has been ongoing since the early 2000s, notably within post-detection **Autonomic[1] Cyber Defense (ACD)** [3]. While ACD also encompasses the preventive (a) and detection (b) stages of cyber defense, we refer to post-detection ACD simply as ACD throughout the remainder of this paper. Various decision-theoretic models utilizing **Artificial Intelligence (AI)** techniques have been proposed (*cf*. Section 3). However, while state-of-the-art approaches acknowledge the dynamic nature of modern IT infrastructures [4], particularly in the era of computer virtualization and service containerization, their impact on security action space has been largely neglected. This oversight mainly stems from limited industry adoption of such decision-theoretic models, caused by factors such as computational complexity [5] and trust [6] constraints. As a result, existing research frequently simplifies security responses into push-button actions executed within controlled test environments. In practice, however, many mitigation actions are more complex, often involving sequences of sub-procedures that rely on specific system or network-level capabilities, which vary depending on the current state of the underlying IT infrastructure.

Therefore, we propose security intent as a promising concept to bridge these gaps. First, intent enables defense systems to maintain a high-level, strategic "understanding" of their objectives before reacting, simulating the reasoning process of human SOC analysts. Second, intent abstracts away intricate implementation details in decision-theoretic mitigation planning, stabilizing its action space and encapsulating the impact of environmental changes (*e.g*., service update, reallocation, *etc*.) of security functions into the non-stationary model parameters of the underlying IT infrastructure. The goal is to facilitate robust handling of increasingly complex and continuously evolving IT infrastructures for autonomic security responses.

---

[1] We upgrade to the term "autonomic" to denote a level of automation between "automated" and "autonomous". Automated systems require human initiation and follow fixed rules; autonomic systems [7] can adapt and self-manage under limited human oversight; autonomous systems operate entirely independently, without human intervention.

In this paper, we demonstrate how ***Intent-Based Networking (IBN)*** [8,9] can be leveraged in ACD to intelligently reconfigure system behavior and security policies in response to active cyber attacks. While IBN was originally developed to simplify network management, we specifically focus on its application for security responses. Building upon this perspective, we seek to answer the following Research Questions:

- **RQ1**. What constitutes a suitable and practical operational definition of security intent in the context of Autonomic Cyber Defense ?
- **RQ2**. How can such security intents be effectively integrated into ACD, using Intent-Based Networking to enable adaptive and high-level response planning?

To address **RQ1**, we propose a unified definition of security response in the form of structured security intents, grounded in the MITRE-D3FEND™ ontology [10]. D3FEND formalizes cyber defense techniques into actionable concepts that can serve as robust building blocks for our intent definition across diverse attack scenarios (see Section 2.2 for an overview of D3FEND). We choose D3FEND over other existing cybersecurity ontologies [11,12] due to its close alignment with the widely adopted MITRE-ATT&CK™ offensive taxonomy [13], which facilitates deeper understanding of cyber attacks (*e.g.*, APTs) and supports precise intent-driven security responses.

To answer **RQ2**, we introduce a general methodology for integrating our proposed security intent representation into state-of-the-art decision-theoretic ACD models. The integration is two-tiered, comprising (1) an ***Intent Discovery Agent (IDA)***, responsible for autonomically generating security intents based on real-time security observations, and (2) an ***Intent Enforcement Agent (IEA)***, tasked with ensuring these intents remain enforced in dynamic operational environments. We further demonstrate a concrete use case showcasing how such intents can be implemented by the IEA beyond traditional networking contexts (originally envisioned by IBN), using the ***Kubernetes project***[2].

The remainder of the paper is organized as follows. Section 2 presents the background, covering IBN and the D3FEND ontology. Section 3 provides a review of related work on both ***Intent-Based Cyber Defense (IBCD)*** and Autonomic Cyber Defense (ACD). Section 4 details our contribution, including the unified intent definition and the two-tiered integration methodology via the IDA and IEA. Finally, Section 5 summarizes our conclusions and outlines future research directions.

## 2    Background

### 2.1    Intent-Based Networking (IBN)

IBN is a powerful paradigm originally developed to simplify computer network configuration by abstracting low-level, device-specific tasks into high-level, device-agnostic intents. According to the ***Internet Engineering Task Force (IETF)*** [8], intent is defined as "operational goals (to be met) or outcomes (to be delivered) defined in a declarative manner without specifying how to achieve them". In the context of post-detection attack mitigation, intent can be further refined to represent a desired system state in which normal functionality has been restored (outcome to be delivered), and the threat impact has been either eliminated or reduced to a level considered tolerable (goal to be met). In most literature on IBN, we consider that an ***Intent-Based Networking System (IBNS)*** periodically receives intent from human users. The intent lifecycle thus has to be separated into two closed loops (see Fig. 1).

---

[2]    Kubernetes is an open-source container orchestration platform that enables dynamic, programmable infrastructure – ideal for evaluating intent enforcement in complex environments.

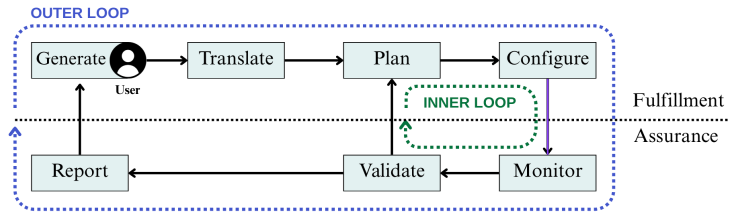

**Fig. 1.** Intent lifecycle (simplified from [8])

a) **Outer closed loop:** involves user interaction. It interprets the user's intent, implements it, and reports the outcome back to the user.
b) **Inner closed loop:** autonomously detects and counteracts intent drift. It does not involve any user in the loop.

Intents may be transient or persistent. Transient intents conclude once their corresponding operation is executed, *i.e.*, they only traverse the outer loop once. In contrast, persistent intents remain active under continuous lifecycle management until explicitly deactivated or removed.

## 2.2    MITRE-D3FEND Ontology

Cybersecurity ontologies can serve as robust building blocks for security intent definition. Among them, the MITRE Corporation's D3FEND ontology [10] provides a structured representation of cybersecurity countermeasures and logically relates them to cyber attacks. D3FEND formalizes cyber defense mechanisms into well-defined, actionable concepts – defensive techniques. Identifying which attacks a countermeasure counteracts is insufficient; it is equally critical to understand, from an engineering standpoint, the concrete mechanisms by which it mitigates those threats [14]. To enhance this understanding, D3FEND is complemented by its offensive counterpart – ATT&CK [13], a conceptual cyber attack taxonomy also developed by the MITRE Corporation, primarily focusing on offensive techniques, which are instantiated by practical attacks.

D3FEND introduces digital artifacts as the basis for conceptualizing and instantiating the relations between offensive techniques (ATT&CK) and defensive techniques (D3FEND) [10]. A digital artifact represents a piece of information or a digital entity that can be acted upon within a cybersecurity context. These attack-defense relationships are covered in D3FEND; let's clarify this with an example: An attacker may establish stealth communication between a compromised server and a remote ***Command and Control (C2)*** server using the ATT&CK technique "Dynamic Resolution", which dynamically resolves malicious domain names to obtain the C2 server's frequently changing IP address. In this scenario, one relevant digital artifact is "Outbound Internet DNS Lookup Traffic", generated by the compromised server as it queries external DNS servers for resolution. From a defensive standpoint, the D3FEND technique "DNS Denylisting" can be applied to block outbound DNS queries to known malicious domains, thereby disrupting the attacker's ability to establish C2 communications.

## 3 State of the art

In this section, we present the state of the art in Intent-Based Cyber Defense (IBCD) and Autonomic Cyber Defense (ACD) focusing on the post-detection phase, which are complementary rather than competing. Specifically, IBCD typically lacks *autonomic intent-generation* capabilities, while ACD requires *high-level action abstraction* to adapt to dynamic IT infrastructures. As illustrated in Fig. 2, these two dimensions can be seen as dual aspects of a unified cyber defense.

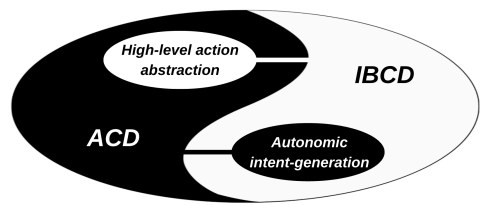

**Fig. 2.** Yin-Yang Relationship Between ACD and IBCD.

### 3.1 Intent-Based Cyber Defense (IBCD)

In this sub-section, we analyze existing pioneering efforts that have sought to combine IBN and cybersecurity in various ways. As summarized in Table 1, for our work, we characterize relevant studies according to the following points:

- **Managing Scope:** The operational domain over which the intent is applied.
- **Intent Enforcement Point:** The network components where the security intent is enforced, ranging from low-level devices to high-level orchestrators.
- **Intent Definition:** refers to how the intent is formally defined within the system. This dimension aligns with the concept of intent expression, as introduced in the taxonomy of intent profiling [9,10]. We go beyond basic expression classification (*e.g.*, template-based) by examining the underlying intent representation.
- **User-specified (U) or Machine-inferred (M):** distinguishes *who* or *what* generates the intent:
  - **U:** The intent is manually provided by human users (*e.g.*, SOC analysts), based on their expertise and organizational policies.
  - **M:** The system can infer intents by interpreting environmental observations (*e.g.*, security alerts), often using AI methods.

**Table 1.** State of the art of Intent-Based Cyber Defense

| Ref. | Managing Scope | Intent Enforcement Point | Intent Definition | U/M |
|---|---|---|---|---|
| [15] | *Network* access control in large/distributed network environments | *Network* routers | Language for ACL (Access Control List) Intents | *U* |
| [16,17] | Cloud-native *network* security services orchestration across virtualized network infrastructures | *Network* security functions (*e.g.*, Firewall, Intrusion Prevention System, *etc.*) | Customer-Facing Interface (I2NSF IETF standard) | *U* |
| [18] | Kubernetes (K8s) *network* isolation in multi-domain and multitenant cloud environments for liquid computing | K8s *network* policy API | Intra/inter-virtual K8s clusters resource communication authorization intent language | *U* |
| [19] | Business applications/Security services orchestration in distributed cloud/edge infrastructures | Business applications (*e.g.*, web server) and security services (e.g., VPN, filtering, *etc.*) | Service deployment policy template (*e.g.*, software/hardware capabilities, latency, *etc.*) | *U/M* |
| [20] | Multilayer secure *network* channel orchestration | Optical *network* controllers; *network* switches, and routers | Encryption centric requirement submitted through API/CLI interface (*e.g.*, in JSON) | *U* |
| [21] | Software Defined *Networking*[3] based Moving Target Defense against reconnaissance attack | Software-Defined *Networking* Controller | Template-based, classification-driven network traffic rerouting policy | *U* |
| [22] | *Network* security services orchestration based on a service mesh framework in cloud-native K8s environments | *Network* security functions | Language for security service policy management | *U* |

From the "Managing scope" column, we observe that most prior works focus predominantly on network-level security. Their defensive schemes affect network traffic possibly involving endpoints[4], but do not directly impact the internal security posture of the endpoints themselves (*e.g.*, processes, accounts, *etc.*). This emphasis is understandable, since IBN is network centric. However, a resilient SOAR platform cannot rely solely on network-level appliances, it must also support endpoint-level capabilities (*e.g.*, filesystem management), which are currently lacking in existing approaches.

Various representations have been proposed to define and manage security intent, each tailored to different management scopes. Notably, some of them blur the line between declarative expression and rule-based control, thereby undermining the level of abstraction that is central to IBN. For instance, the authors of [16,17] introduced an intent-based ***Network Security Functions (NSF)*** orchestration system using the Interface to NSF framework (I2NSF) [23]. An NSF inspects and optionally modifies packets traversing networks, such as a firewall and a network intrusion detection system. In

---

[3] Software Defined Networking separates the network control plane from the data plane to enable centralized, programmable network management.

[4] In this paper, a network endpoint refers to any physical or virtual entity that communicates over a network, including servers, containerized applications and so on.

[17], a customer-facing interface based on YANG data model[5] was employed for user intent acquisition. However, this interface supports an event-condition-action scheme, which is typically considered rule-based. Similarly, [15] proposed a language for network *Access Control List (ACL)*[6] intents to manage router-level packet forwarding policies based on IP prefixes. In this model, users can either directly provide explicit ACL rule updates, or express high-level reachability goals, representing security intents.

Most existing approaches only support user-specified intents. While [19] briefly mentioned the possibility of machine-generated intents, they did not provide further elaboration. As a result, the current state of the art of Intent-based Cyber Defense largely overlooks the possibility of machine-inferred intents. This omission neglects the abstraction power that intent-based models offer, which would significantly simplify the underlying decision-making problem, *i.e.,* selecting the most promising security intent given the current observed context.

### 3.2    Autonomic Cyber Defense (ACD)

In the context of ACD, decision-making is inherently essential. Recent research has explored various decision-theoretic approaches, leveraging *Game Theory* and *Machine Learning* [3]. For instance, the authors of [24] investigated the application of model-free Tabular Q-Learning, a type of *Reinforcement Learning*, applied to a Markov Decision Process to generate a defense strategy (*i.e.*, a mapping from estimated circumstances to actions that guide the defender's decisions) offline[7] within a simulated network environment. At each time step, the defender selects a high-level action such as "restore a host into a known good state", based on uncertain knowledge of the network's current compromise and availability state. Though these actions resemble intent-based abstractions, they emerge not from an explicitly intent-driven model but from the simplification of the simulation environment and its attack surface.

Popular black-box Machine Learning techniques are powerful but lack explainability: they do not explicitly justify why a particular action was taken, relying solely on empirical patterns learned from data. This limitation hinders their integration into human-supervised SOAR platforms. While post hoc explainability methods such as SHAP [25] can attribute feature importance (*i.e.*, the contribution of each input feature to a specific model prediction), they fall short of providing decision-level explanations that are essential for operational trust and oversight by human experts.

In contrast, [26] treats ACD directly as a Partially Observable Markov Decision Process. Their decision-theoretic planner relies on a risk-aware Cyber Security Game simulator [27], which abstracts attacker-defender interactions across coarse-grained infrastructure models, enabling optimal defense action planning based on the inferred risk expectation. More recently, the authors of [4] proposed *Conjectural Online Learning*

---

[5]    YANG is a hierarchical, schema-based language typically for network management protocols.
[6]    A network ACL is a set of rules applied to network interfaces, typically on routers and firewalls, to permit or deny traffic based on criteria such as IP address, protocol, or port number.
[7]    Offline learning refers to training the Reinforcement Learning agent entirely during development phase, without further strategy updates during deployment phase.

*(COL)*, a model-based Game Theoretic planner that treats the attacker-defender inter-action as a non-stationary, partially observed stochastic game. In COL, each player maintains a dynamic conjecture on both the underlying game parameters and the oppo-nent's strategy, both of which may be misspecified. According to the authors, this is the first security application to simultaneously handle learning under misspecified mod-els and bounded computational resources [4]. As a result, COL naturally adapts to changing infrastructures and yields explainable action choices.

Nevertheless, mitigation actions employed by state-of-the-art ACD systems remain simplistic. They typically incorporate straightforward actions, either overlooking prac-tical implementation complexity or relying on a predetermined implementation method, suitable only for simulated environments. However, real-world IT infrastruc-tures typically involve multiple tools for overlapping purposes, each necessitating ven-dor-specific configurations and possessing distinct advantages and risks. Hence, alt-hough [4] assumes non-stationary IT infrastructures, the fact that it still operates over a fixed action alphabet is problematic. Only an intent-based layer, that decouples high-level security goals from concrete, vendor-specific configurations, can make plan-ners like COL practical across diverse and evolving IT infrastructures.

## 4 Proposed Contribution

We integrate Autonomic Cyber Defense (ACD) and Intent-Based Cyber Defense (IBCD) into an ***Intent-Based Autonomic Cyber Defense*** framework, which can be viewed either as an ACD enhanced with *high-level action abstraction*, or as an IBCD capable of *autonomic intent-generation* (*cf*. Fig. 2). Adopting the latter Intent-Based Networking (IBN) perspective within an intent lifecycle representation (*cf*. Fig. 1), we designate the outer loop's autonomic intent generator as the ***Intent Discovery Agent (IDA)***, and the underlying Intent-Based Networking System (IBNS) – which consumes each "discovered intent" and implements the remaining components of the lifecycle – as the ***Intent Enforcement Agent (IEA)***. Before detailing these two agents, we first introduce our security intent definition.

### 4.1 Unified Security Intent

Security observations typically consist of either standalone alerts or incidents, *i.e.*, a chain of correlated alerts. An alert $AL$ is defined as structured attack-related data (*e.g.*, in JSON) conforming to a well-defined schema depending on the detection software. We suppose that each alert is mapped to one ATT&CK technique, which is commonly supported in modern intrusion detection solutions.

We propose a unified security intent definition based on the D3FEND ontology. Formally, facing an alert $AL$, a security intent $I_{AL}$ is defined by the following tuple:

$$I_{AL} \triangleq \langle OT, MD, DA_T, DT \rangle$$

Where $OT$ is the identified offensive technique employed by $AL$. $MD$ is the alert technical metadata *s.t.* $MD \subseteq AL$, excluding non-technical fields irrelevant to intent ful-fillment (*e.g.*, alert ID). $DT$ is a defensive technique. $DA_T$ is the digital artifact targeted both by $OT$ and $DT$ for offensive and defensive purposes, respectively. Intuitively, the goal of $I_{AL}$ is to prescribe a $DT$ that operates on a specific $DA_T$ targeted by $OT$, thereby reaching a desired state where the security risks identified by $AL$ are mitigated. The intent $I_{AL}$ embodies a minimal response strategy, aiming to apply the least intrusive yet effective countermeasure. We elaborate this after extending D3FEND for our use case.

**A Lightweight Extension of D3FEND for Operational Use**

D3FEND exhibits object-oriented features and is based on the Web Ontology Language [28]. Offensive and defensive techniques both rely on digital artifacts to establish their presence on target infrastructure, which comprises both digital and physical artifacts[8]. This dependency is characterized through *properties*, drawn from two partially overlapping[9] property spaces – $Props_O$ and $Props_D$, corresponding to offensive and defensive semantics, respectively. We denote the dependency relation by $OT \xrightarrow{Prop} DA$, or $OT.Prop = DA$, where $Prop \in Props_O \cup Props_D$.

In D3FEND, the functions of an offensive or defensive technique can be represented as a set of property restrictions over digital artifacts. An offensive technique $OT$'s restriction on property $Prop_O$ over digital artifact $DA$ is denoted as $Restrict_{OT}(Prop_O, DA)$, and is defined as an anonymous superclass of $OT$ that enforces the relation $OT \xrightarrow{Prop_O} DA$.

We propose a categorization of offensive properties based on their semantic roles, as summarized in Table 2. Each offensive property $Prop_O$ belongs to one category. An offensive property can be prefixed with *may* in a restriction $Restrict_{OT}(may\text{-}Prop_O, DA)$. This denotes that an attack instance of $OT$ does not require the engagement $OT \xrightarrow{Prop_O} DA$ to succeed, *i.e.*, that restriction is *optional* at the $OT$ class level, and can be omitted by its instances. However, if an attack instance of $OT$ does engage $DA$ via $Prop_O$, its corresponding restriction becomes *valid* (or mandatory) and the "*may-*" qualifier of $Prop_O$ is dropped for clarity. Any restriction of a given offensive technique without prefixed offensive property is by default valid for all its attack instances.

**Table 2.** Categorization of Properties linking Offensive Techniques and Digital Artifacts

| Category | Definition | Property Instances |
|----------|-----------|-------------------|
| Alter | Modifies existing digital artifacts | *obfuscates, encrypts, may-encrypt, …* |
| Generate | Creates new digital artifacts | *forges, produces, may-produces, …* |
| Exploit | Leverages existing digital artifacts | *invokes, injects, may-injects, …* |
| Remove | Removes existing digital artifacts | *deletes, erases, may-erases, …* |

On the defensive side, each property $Prop_D$ belongs to one of the categories defined as defensive tactics [10]. We focus on intents derived from incident responding tactics (*cf.* Table 3). A defensive technique's restriction $Restrict_{DT}(Prop_D, DA)$ denotes its intended capability to manage $DA$ in accordance with the semantics of $Prop_D$.

---

[8] While attackers might physically infiltrate the target infrastructure, the scope of our defensive techniques is strictly limited to digital artifacts hosted on physical resources.

[9] For instance, *encrypts* exists as both offensive and defensive property with distinct goals.

**Table 3.** Categorization of Properties linking Defensive Techniques and Digital Artifacts

| Category | Definition | Property Instances |
|---|---|---|
| Evict | Removes digital artifacts through which attackers establish their presence on target systems | *terminates, disables, ...* |
| Isolate | Creates barriers that prevent adversary access to digital artifacts. | *blocks, restricts, ...* |
| Restore | Restores digital artifacts to a known better state | *restores* |

**Minimal Security Response**

Valid restrictions associated with a given attack instance of an offensive technique form a conjunction. Therefore, a minimal security response that disrupts the impact on any engaged digital artifact can invalidate the offensive technique of that attack. We favor minimal security responses because post-detection Autonomic Cyber Defense typically rely on sequential decision-making (see Section 4.2). Keeping individual responses minimal not only reduces collateral impact but also preserves strategic flexibility.

Suppose an attack instance has valid restriction $Restrict_{OT}(Prop_O, DA)$; a defensive technique $DT$ **may** invalidate $OT$ (and thus the attack) through $DA$ if and only if there exists a defensive property $Prop_D$ *compatible* with $Prop_O$ s.t. $OT \xrightarrow{Prop_O} DA \xleftarrow{Prop_D} DT$. The pair $(Prop_O, Prop_D)$ are considered compatible if and only if their respective categories are compatible, as defined in Table 4. The compatibility constraint prevents invalid countermeasures – *e.g.*, *restore* a "malicious connection" *produced* by an attacker.

It is important to emphasize the uncertainty inherent in the above formulation. There is no guarantee that even a compatible $Prop_D$ can mitigate the effect of $Prop_O$ on the same digital artifact – its success depends on implementation details and the evolving operational environment. Moreover, from the defender's perspective, identifying valid restrictions associated with a suspected attack is a non-trivial task and not sufficient either: the engaged digital artifacts must be instantiated with concrete attributes (*e.g.*, IP addresses for Network Traffic artifacts) to provide actionable information for the defensive technique to operate on. D3FEND, by design, does not define such attributes, as they depend on the specifics of the underlying IT infrastructure. We argue that the representation of digital artifact attributes must therefore follow use-case requirements.

**Table 4.** Compatibility of Defensive vs. Offensive Property Categories

| *Defensive* \ *Offensive* | Alter | Generate | Exploit | Remove |
|---|---|---|---|---|
| Evict | ✓ | ✓ | ✓ | ✗ |
| Isolate | ✓ | ✓ | ✓ | ✗ |
| Restore | ✓ | ✗ | ✓ | ✓ |

Let $DA.Prop_D^{-1}$ be the set of all defensive techniques $DTs$ s.t. $\forall DT \in DTs$: $DT.Prop_D = DA$. Given a defensive technique $DT$ whose set of property-artifact restrictions is exactly $\{Restrict_{DT}(Prop_i, DA_i) | i \in [1, N]\}$, we also define $DT$'s equivalence class $\{DT\} \triangleq \bigcap_{i=1}^{N} DA_i.Prop_i^{-1}$. Defensive techniques belonging to one equivalence class pursue the same defensive objective, but differ in their implementation strategies – just as multiple policies supporting the realization of a single intent.

We illustrate the notion of minimal response through a concrete example (*cf.* Fig. 3). Pluggable Authentication Modules (PAM) handle authentication across various services in Unix-like systems. Attackers can exploit PAM components to intercept credentials or grant unauthorized access. A PAM-based attack invariably modifies *Authentication Service*, in cases where it additionally requires modifying *OS Configuration File*, any of the four defensive techniques (highlighted in blue), if successfully executed, is sufficient on its own to mitigate the attack. Security experts would likely rank their effectiveness[10] as: *System Call Filtering* (most effective) > *Process Termination* > *File Eviction* > *Host Reboot*. While *Host Reboot* and *Process Termination* belong to the same equivalence class, they differ in operational cost and mitigation impact – particularly depending on whether the attacker has established persistence on the host.

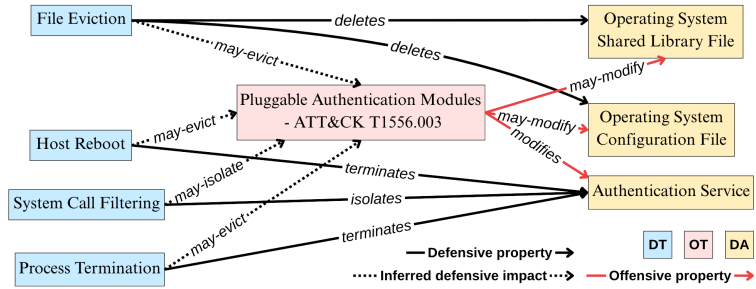

**Fig. 3.** Truncated representation of ATT&CK Technique T1556.003 in D3FEND.

## 4.2 Methodology for Intent Discovery Agent (IDA)

The IDA is responsible for autonomic mitigation intent generation. While numerous models fall under the hood of decision theory, the current state-of-the-art models in Autonomic Cyber Defense (*cf.* Section 3.2) fundamentally address *sequential strategic decision-making under uncertainty* [3,29]. This uncertainty arises naturally from the defender's partial knowledge of the operational environment and ongoing attacks. Sequential decision-making is essential, as each attacker-defender interaction yields new insights into the environmental dynamics and adversary behavior, requiring defender to continuously update its strategy. A standalone or static decision-making approach would fail to benefit from such observations, thus risking to become obsolete facing evolving attacks. Furthermore, maintaining a dynamic strategy provides continuous transparency for human supervision across the entire defense lifecycle, supporting oversight under a wide range of possible circumstances.

### Preliminary: Standard Action-based ACD Modeling Methodology

Before introducing the methodology for our IDA, we present a standard action-based ACD model that aligns with the requirements outlined above. State-of-the-art approaches, as explained in [29], typically rely on ***Partially Observable Markov Decision Process (POMDP)*** – modeling only the defender, or ***Partially Observable Stochastic***

---

[10] Here, effectiveness reflects a multi-criteria tradeoff, including success probability, operational cost, response time and potentially more context-dependent metrics.

***Game (POSG)*** – modeling both attacker and defender. Note that a POMPD can be viewed as a single-agent POSG [29], and is defined by the following tuple:

$$\Gamma \triangleq \langle \mathcal{S}, \mathcal{A}, \mathcal{O}, t, r, z, b_1, \gamma \rangle$$

Where $\mathcal{S}$ is the state space (*e.g.,* a state could be the number of compromised servers); $\mathcal{A}$ is the action space (*e.g.,* an action might be a playbook for endpoint isolation); and $\mathcal{O}$ is the observation space (*e.g.,* outputs from intrusion detectors). In our case, each observation $o \in \mathcal{O}$ comprises alerts paired with their associated ATT&CK offensive technique. The state transition function $t(s'|s, a)$ denotes the probability of reaching state $s'$ from $s$ by taking action $a$. The reward function $r(s, a)$ assigns a scalar utility to action $a$ in state $s$, capturing the net security benefit by weighting risk mitigation – across security goals such as availability, confidentiality, and integrity – against operational cost. The observation function $z(o|s)$ denotes the probability of observing $o$ in state $s$. The real number $\gamma \in [0,1]$ is the long-term reward discounting factor. $b_1$ is the initial belief state – In POMDP, the true system state is not directly observable. Instead, the defender maintains a belief state – a probability distribution over the state space $\mathcal{S}$ that captures the current estimate of the environment based on past actions and observations. The defender's goal is to find an optimal deterministic strategy $\pi^* : \mathcal{B} \to \mathcal{A}$ (where $\mathcal{B}$ is the belief space) that maximizes the expected ($\mathbb{E}$) discounted long-term reward (where time $t$ is assumed to be discrete):

$$\pi^* = \arg\max_\pi \mathbb{E}_\pi \left[\!\left[ \sum_{t=1}^{\infty} \gamma^{t-1} \cdot r(s_t, a_t) \mid b_1 \right]\!\right]$$

Computing $\pi^*$ typically involves a range of techniques (*e.g.*, Reinforcement Learning). However, algorithmic details are beyond the scope of this paper. We assume the existence of an optimal strategy [30] within the proposed model applied to the cyber defense.

**IDA Methodology**

The concept of intent surpasses that of one-time action due to its potential durability. Suppose an attacker has stealthily controlled a network switch before compromising a connected server. Even if the defender subsequently isolates the server by restricting traffic on adjacent switches (including the compromised one), the attacker may still regain access by reestablishing an unauthorized communication path through the infiltrated switch. To mitigate such risks, persistent intents are monitored by the IEA (see Section 4.3), ensuring that they remain consistently enforced. We therefore propose an intent-based extension of $\Gamma$ inspired and supported by classic IBNSs [9]:

$$\Gamma' \triangleq \langle \boldsymbol{\mathcal{S}'}, \boldsymbol{\mathcal{A}'}, \mathcal{O}, t, r, z, b_1, \gamma, \boldsymbol{\lambda} \rangle$$

Where the state space $\mathcal{S}' = \mathcal{S} \times \mathcal{I}$ with $\mathcal{I}$ denoting the persistent intent store space. An Intent Store $IS = \{(I_1, TTL_1), (I_2, TTL_2), \dots\} \in \mathcal{I}$ is a collection of active persistent intents $I \in \mathbb{I}$ (the intent space), each associated with a *Time To Live (TTL)* value. After each round, the $TTL$ of every intent in $IS$ is decremented by 1. Intents with $TTL = 0$ are purged from the store at the beginning of each round. The defender maintains an Intent Store to enforce long-term mitigations that remain effective across multiple decision epochs, ensuring resilience against persistent or recurring attacks.

The additional intent observation function $\lambda : \mathcal{O} \rightarrow \mathbb{I}$ maps an observation $o$ to a candidate intent set $\mathbb{I}_{cand}$[11]. It constrains the action space based on the current observation, thereby reducing planning complexity and minimizing disruption to normal operations. For each alert-technique pair $(AL, OT) \in o$, the function $\lambda$:

1. instantiates (via a rule-based or Machine-Learning-based mapper) all digital artifacts instances engaged by $OT$, using the alert technical metadata $MD \subseteq AL$. Then,
2. identifies, through queries over the D3FEND ontology (*e.g.*, using SPARQL), all defensive techniques that **may** invalidate $OT$ through any operable (sufficiently instantiated) digital artifact $DA_T$, as defined in Section 4.1.

Combining the results of each iteration we get:

$$\mathbb{I}_{cand} = \{\langle OT, MD, DA_T^i, DT_i \rangle \mid \forall (AL, OT) \in o, MD \subseteq AL, \exists i:$$
$$OT \xrightarrow{Prop_O^i} DA_T^i \xleftarrow{Prop_D^i} DT_i, \text{ where } \left(Prop_O^i, Prop_D^i\right) \text{ are compatible and } DA_T^i \text{ is operable}\}$$

If the security observation lacks sufficient information (depending on the specific use case) to instantiate a digital artifact instance exploited by the offensive technique, then the corresponding artifact will be omitted by $\lambda$. This also applies to cases where that property-artifact restriction is optional and not exploited by the suspected attack.

The action space $\mathcal{A}'$ comprises four action types: **(a)** insert a new intent $I \in \mathbb{I}_{cand}$ as persistent[12] to $IS$, **(b)** modify a persistent intent of $IS$, **(c)** execute an intent $I \in \mathbb{I}_{cand}$ as transient (at lower cost), and **(d)** take no action. For each type except (d), the affected intent's $TTL$ is (re)set to a default value. For (b), the replacing intent must preserve the offensive context $(OT, MD)$ while updating the defensive measure $(DA_T, DT)$.

Notably, feedback from the underlying IBNS, which identifies cases of misimplemented intents, should be incorporated into the reward function $r$. This allows the defender to account for the actual implementation status of intents in the Intent Store. For example, *File Eviction* technique can fail due to permission issues, particularly in cases where the attacker has gained control over filesystem privileges without being detected.

We illustrate the proposed methodology in a concrete use case (*cf.* Fig. 4). Suppose an attacker has established persistence on a *Linux* server by creating a scheduled job `maljob` that periodically runs a malicious script `malicious.sh`. This script resolves the DNS name `c2.malicious.com` – whose underlying IP rotates, and connects to the resolved address to maintain its Command-and-Control (C2) channel.

An IDA based on our proposed intent-based model automatically derives intents and can be trained, for instance using Adversarial and/or or Reinforcement Learning [3,29], to learn an optimal mitigation strategy. In our case, when both *Alert-Technique 1* and *Alert-Technique 2* are present in a security observation, the digital artifacts engaged by the ongoing attack are instantiated (highlighted in yellow). The IDA may prioritize blocking the malicious connection between the compromised host and the attacker's C2 server (*Persistent Intent 1*), before evicting the malicious content planted on the host (*Persistent Intent 2*). These two persistent intents are sent to the underlying IEA for intent enforcement, which is detailed in the next section.

---

[11] As stated earlier, we focus on post-detection responses: the defender only acts upon valid security observations. Otherwise, no action is taken – preventive measures are excluded.

[12] The persistent or transient property of an intent is considered extrinsic.

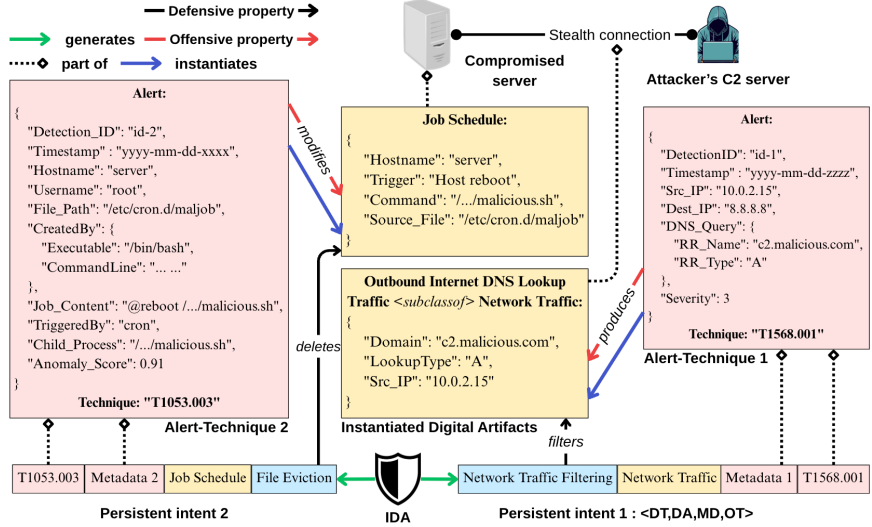

**Fig. 4.** A concrete use case of intent-based attack mitigation for a persistent attack

### 4.3 Methodology for Intent Enforcement Agent (IEA)

The IEA receives security intents generated by the IDA and is responsible for their enforcement. It behaves as a typical IBNS, encompassing the full intent lifecycle except intent generation. However, due to the wide range of defensive techniques represented in the D3FEND ontology, the IEA must be highly versatile. To our knowledge, the most versatile existing framework for Intent-Based Cyber Defense – *i.e.*, capable of supporting a broad spectrum of defensive techniques – is the ***Intent-based Closed-loop Security Control (ICSC)*** framework proposed in [17], which builds upon the IETF standard Interface to Network Security Functions (I2NSF) framework (*cf.* Section 3.1).

Nevertheless, ICSC is inherently limited to network traffic based security, as it relies exclusively on Network Security Functions (NSFs). The IEA relaxes this constraint by generalizing the framework to include ***Endpoint Security Functions (ESFs)*** as well. An ESF, as opposed to an NSF, enforces control within the internal environment of a network endpoint, regardless of whether the enforcement is executed locally or orchestrated remotely[13]. We thus reinterpret the five core components of the ICSC framework as the combination of an IDA and an IEA, as follows (*cf.* Fig. 5):

- **I2NSF User:** corresponds to our IDA.
- **Our proposed IEA Architecture:**
    - ***Security Functions (SFs):*** Enforcement points responsible for executing network-level (NSF) or endpoint-level (ESF) security responses.
    - ***Security Controller:*** The central orchestrator that receives intents and invokes appropriate security functions. It translates intents into low-level configurations based on the advertised capabilities of available SFs (see next line).

---

[13] For instance, user account management may be delegated to external authorities (*e.g.*, Azure Active Directory), which still exert control over internal endpoint behavior.

- **Developer's Management System:** provides a platform for developers (or administrators) to register and advertise SFs' capabilities. An SF's capability indicates the specific defensive technique(s) that it can realize, along with the operational parameters required for execution.
- **Analyzer:** ensures intent assurance (*cf.* Fig. 1). In [17], the Analyzer is also in charge of collecting alerts, but here this role is well assumed by the IDA.

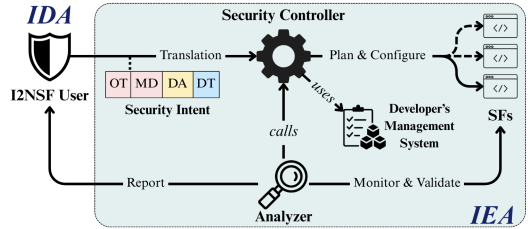

**Fig. 5.** An adapted overview of the ICSC framework conforming to the intent life cycle

We validate our proposal by implementing the two intents derived in the previous use case (*cf.* Fig. 4). Noting that *Network Traffic Filtering* is network-level and *File Eviction* is endpoint-level, we adopt the Kubernetes (K8s) platform, which provides both NSF and ESF capabilities. Our testbed consists of a K8s cluster composed of one controller and several Pods (the smallest deployable application endpoints in K8s). We apply both mitigation techniques to a single infected Pod and demonstrate their effectiveness. The implementation details and corresponding code are available here[14].

## 5 Discussion and Conclusion

In this paper, we envision a next-generation Security Orchestration, Automation, and Response platform in which the role of security analysts gradually shifts from direct decision-making to supervisory oversight. To this end, we bridged two emerging but independently studied paradigms – Intent-Based Cyber Defense and Autonomic Cyber Defense – by introducing a unified, ontology-driven representation of security response intents, along with a two-tiered methodology for their integration into decision-theoretic Autonomic Cyber Defense models. The integration comprises an Intent Discovery Agent and an Intent Enforcement Agent, which together form a complete intent lifecycle. We illustrated the implementation of the proposed intents through the Intent Enforcement Agent in a Kubernetes-based environment. While this work lays the conceptual and architectural foundation for integrating intent into post-detection Autonomic Cyber Defense systems, we did not evaluate the performance of our proposed methodology. Future work will focus on benchmarking the performance of state-of-the-art Autonomic Cyber Defense systems boosted by Intent-Based Networking, thereby enabling a more comprehensive evaluation across complex infrastructures and dynamic attacks.

**Acknowledgements.** This work was supported by the French National Research Agency (ANR) under the ANCILE project (Grant No. ANR-23-CE39-0010).

---

[14] https://github.com/Zequan99/EDOC-k8s-DT-implementation/tree/master